\documentclass[aps, prd, twocolumn, lengthcheck, superscriptaddress, 
nofootinbib]{revtex4-1}

\usepackage{epsfig}
\usepackage[usenames]{color}
\usepackage{graphicx}
\usepackage{amsmath}
\usepackage{multirow}
\usepackage{epstopdf}

\newcommand\sect[1]{\emph{#1.}---}
\def\bi{\bibitem}

\def\la{\langle}\def\ra{\rangle}
\def\be{\begin{eqnarray}}\def\ee{\end{eqnarray}}
\def\lsim{\mathrel{\rlap{\lower3pt\hbox{\hskip1pt$\sim$}}
     \raise1pt\hbox{$<$}}} 
\def\gsim{\mathrel{\rlap{\lower3pt\hbox{\hskip1pt$\sim$}}
     \raise1pt\hbox{$>$}}} 
\def\del{\partial}

\allowdisplaybreaks

\begin{document}

\title{What's in the core of massive neutron stars ?}

\author{Yong-Liang Ma}
\email{ylma@ucas.ac.cn}
\affiliation{School of Fundamental Physics and Mathematical Sciences,
Hangzhou Institute for Advanced Study, UCAS, Hangzhou, 310024, China}
\affiliation{International Centre for Theoretical Physics Asia-Pacific, Beijing/Hangzhou, China }
\affiliation{Center for Theoretical Physics and College of Physics, Jilin University, Changchun, 130012, China}

\author{Mannque Rho}
\email{mannque.rho@ipht.fr}
\affiliation{Universit\'e Paris-Saclay, CNRS, CEA, Institut de Physique Th\'eorique, 91191, Gif-sur-Yvette, France }

\date{\today}

\begin{abstract}
When hadron-quark continuity is formulated in terms of a topology change at a density higher than twice the nuclear matter densiy $n_0$ the core of massive compact stars can be described in terms of  quasiparticles of fractional baryon charges, behaving neither like pure baryons nor deconfined quarks. Hidden symmetries, both local gauge and pseudo-conformal (or broken scale),  emerge and give rise to the long-standing quenched $g_A$ in nuclear Gamow-Teller transitions at $\sim n_0$ and to the pseudo-conformal sound velocity $v_{pcs}^2/c^2\approx 1/3$ at $\gsim 3n_0$. These properties are confronted with the recent observations in superallowed Gamow-Teller transitions and in astrophysical observations.
\end{abstract}

\maketitle

\sect{ Introduction}
The phase structure of the strong interactions at high densities has been investigated for several decades but still remain more or less totally uncharted. The observation of massive neutron stars and detection of gravitational waves from neutron star merger provide indirect information of nuclear matter at low temperature and high density, say,  up to  ten times the saturation density $n_0$. So far, such phenomena can be accessed by neither terrestrial experiments nor lattice simulation.

The study of dense matter in the literature has largely relied on either phenomenological approaches anchored on density functionals or effective field theoretical models implemented with certain QCD symmetries, constructed in terms of set of relevant degrees of freedom appropriate for the cutoff chosen for the effective field theory (EFT),  such as  baryons and  pions, and  with~\cite{baymetal,quarkyonic} or without~\cite{Holt:2014hma}  hybridization with quarks, including other massive degrees of freedom.  
The astrophysical observations indicate that the density probed in the interior of neutron stars could be as high as $\sim 10$ times the normal nuclear matter density $n_0\simeq 0.16$ fm$^{-3}$ and immediately raise the question as to what the interior of the star could consist of, say, baryons and/or quarks  and a combination thereof. Asymptotic freedom of QCD implies that at some superrhigh  density, the matter could very well be populated by deconfined quarks~\cite{collins-perry}. But the density of the interior of stars is far from the asymptotic and hence perturbative QCD cannot be reliable there. Lacking lattice QCD,  high density at low temperature cannot be theoretically accessed.

In this Letter, a conceptually novel approach going beyond the standard chiral EFT (denoted as $s\chi$EFT) to higher densities $n\gg n_0$ is formulated and is used to predict that the core of massive compact stars constitutes of {\it confined} quasi-fermions of fractional baryon charge, encoding the equation of state (EOS) with  the ``pseudo-confomal (PC)" sound velocity (SV) $v_{pcs}^2/c^2\approx 1/3$.  We suggest that this phenomenon, together with the ``quenched $g_A$ problem" in nuclei, certain hidden symmetries, emerge in nuclear dynamics.

\sect{Emergence of Hidden Symmetries}
The key ingredients in generating higher-energy scales for going beyond the  $s\chi$EFT --- that contains only baryons and (pseudo-)Nambu-Goldstone (NG) bosons --- are two symmetries that are invisible in the vacuum of  QCD: Hidden gauge symmetry and hidden scale symmetry. Both of them could be generated at low density, at higher-order expansions in $s\chi$EFT.  But they must be constrained at high densities. This is because among others the ultraviolet completion of the effective theory involved is  lacking. Our approach is to exploit the possible emergence of these symmetries as density increases to the regime relevant to compact stars, say, $\lsim 10 n_0$.

The higher-energy degrees of freedom we will focus on are the lowest-lying vector mesons $V=(\rho,\omega)$ and the scalar $f_0(500)$.

For the vectors $V$, we adopt the strategy of hidden local symmetry (HLS)~\cite{yamawaki} which at low density is gauge equivalent to nonlinear sigma model, the basis of $s\chi$EFT. This is for two reasons. First, at higher densities, we can assume HLS satisfies Suzuki's theorem~\cite{suzuki}  that states ``when a gauge-invariant local field theory is written in terms of matter fields alone, a composite gauge boson or bosons must be inevitably formed dynamically."  Secondly, the HLS fields could be (Seiberg-)dual to the gluons~\cite{Komargodski,kanetal,karasik}. This first accounts for the presence of the ``vector manifestation (VM)"~\cite{VM} that the $V$ meson mass goes to zero as the gauge coupling $g_V$ flows to zero at some theoretically unknown high density $n_{vm}$. We will see below that $n_{vm}\gsim 25 n_0$ is indicated for the emergence of the PC velocity in stars. Second, it can account for why HLS at the leading power counting works so well and indicate a Higgs phase-topological phase transition coinciding with the quark deconfinement most likely irrelevant to the compact stars we are concerned with~\cite{D}.

Now as for the scalar $f_0(500)$ that we will identify with the dilaton $\chi$ as the NG boson of scale symmetry, we adopt the ``genuine dilaton" (GD) structure proposed in~\cite{crewther}. The key premise of this idea is the existence of an infrared fixed point (IRFP) with the beta function $\beta (\alpha_{s}^{IR})=0$ for flavor number $N_f=3$. The characteristic feature of this IRFP which plays a crucial role in our development is that in the chiral limit both chiral and scale symmetries are in the NG mode. Given the proximity in free-space mass between the kaon and $f_0$, the s(trangeness) flavor figures together with the u(p) and d(own) flavors. In dense medium, however, the effective mass of the s-quark-baryons is to drop less importantly than the light (u,d) quark baryons, thus projecting the GD idea to the two flavors ignoring strangeness as we do in what follows could be justified.

We will confine ourselves as feasible as possible to the main line in our arguments, referring for details to~\cite{MR-review}.

We exploit two versions of the same effective Lagrangian, one purely bosonic incorporating the pseudo-scalar NG bosons $\pi\in SU(3)$, the dilaton $\chi$ and the vectors $V$ in consistency with the symmetries concerned.  We denote the resulting scale-symmetric HLS Lagrangian as $\chi$mHLS with ``m" standing for the meson fields. To do nuclear physics with it, baryons are generated as skyrmions with $\chi$HLS.
And the second is to introduce baryon fields explicitly into $\chi$HLS, coupled scale-hidden-local symmetrically to $\pi$ and $\chi$. This Lagrangian will be denoted as $\chi$bHLS with ``b" standing for baryons.

\sect{Hadron-Quark Continuity and Topology Change}

Given the two Lagrangians, how does one go about doing nuclear many-body problem ?

While in principle feasible, there is, up to date, no simple way to systematically and reliably formulate nuclear many-body dynamics in terms of skyrmions with $\chi$mHLS. We will therefore resort to $\chi$bHLS. With $\chi$bHLS whose parameters are suitably fixed in medium matched to QCD correlators~\cite{VM,MR-review}, we resort to formulating many-body nuclear problems by a Wilsonian renormalization-group (RG) type approach to arrive at Landau(-Migdal) Fermi-liquid theory. The mean field approximation with $\chi$bHLS can be identified with Landau Fermi-liquid fixed point (FLFP) theory valid in the limit $1/\bar{N}\to 0$ where $\bar{N}=k_F/(\Lambda-k_F)$ (with $\Lambda$ the cutoff on top of the Fermi sea)~\cite{shankar,MR91}. This approach can be  taken as a generalization of the energy-density functional theory familiar in nuclear physics~\cite{MR-review}. One can go beyond the FLFP in  $V_{lowk}$-RG as we will do in numerical calculations.

As will be briefly summarized below, the $\chi$bHLS applied to nuclear matter --- call it GenEFT --- is found to describe nuclear matter at $n\sim n_0$ as well as the currently successful $s\chi$EFT to typically N$^4$LO. There is a however a good reason to believe that the $s\chi$EFT with the Fermi momentum $k_F$ taken as a small expansion parameter must necessarily breakdown at high densities relevant to massive compact stars.  The appearance of quark degrees of freedom in one form or other is a natural candidate for the breakdown mechanism.  In our approach, where and how this breakdown occurs can be determined by the skyrmion approach with the bosonic Lagrangian $\chi$mHLS. This is because topology in $\chi$mHLS can be traded in for  what corresponds to hadron-to-quark continuity.

The skyrmion approach, while too daunting to handle nuclear dynamics directly and systematically, can however provide  valuable and robust information on the possible topological structure involved in going beyond the normal nuclear matter density regime. At some density above $n_0$ --- denoted as $n_{1/2}$ --- the matter described in terms of skyrmions is found to transform to that of half-skyrmions.
This transition has several highly correlated important impacts on the EOS for densities $n > n_{1/2}$. The most crucial of them are that~\cite{MR-review} at $n_{1/2}$ (a) the quark condensate $\la\bar{q}q\ra$ vanishes {\it globally} but not locally with non-vanishing pion and dilaton decay constants ($f_\pi\sim f_\chi \neq 0)$, resembling the pseudo-gap phenomenon in condensed matter,   (b) a parity-doubling emerges in the baryon spectrum with the chiral invariant mass $m_0 \simeq (0.6-0.9)m_N$ and (c) the hidden gauge coupling associated with the $\rho$ meson coupling starts to drop and flows to zero at the VM fixed point~\cite{VM}, hence the vector meson becoming massless. This is in line with the Suzuki theorem~\cite{suzuki}.

Our key strategy is to incorporate these effects connected with the topology change in $\chi$mHLS into the parameters of the $\chi$bHLS  in formulating GenEFT.

It is interesting to note that the skyrmion to half-skyrmion transition in the skyrmion matter has an analog in condensed matter in (2+1) dimensions. For instance, the transition from the magnetic N\'eel ground state to the VBS (valence bond solid) quantum paramagnet phase~\cite{DQCP}.  The half-skyrmions intervening here are {\it deconfined}, thus the transition involves ``deconfined quantum critical points" but with no local order parameters for the phase transition.  What takes place in dense skyrmion matter seems however quite different because the half-skyrmions are not deconfined but confined by monopoles~\cite{cho}. This suggests that the confined half-skyrmion complex be treated as a local baryon number-1 field with its parameters, such as the mass, coupling constants etc. drastically modified from the vacuum quantities reflecting the topology of the half-skyrmions. This allows us to work with a single Lagrangian $\chi$bHLS with the parameters controlled  by topology before and after the topology change density $n_{1/2}$. We suggest $n_{1/2}$ to be the density at which the hadron-quark change-over takes place. There is no phase transition involved.

\sect{Quenched $g_A$ in Nuclei and Continuity in Scale Symmetry}
As detailed in \cite{MR-review}, the GnEFT so formulated works fairly well, in fact as well as the s$\chi$EFT up to, say, N$^4$LO, for baryonic matter properties for $n_0$ and perhaps even up to $\sim 2n_0$. With the heavy vector and scalar degrees of freedom explicitly figuring in the dynamics, the power counting rule is considerably different from that of s$\chi$EFT in which the vector and scalar excitations enter at loop orders. The mean field with $\chi$bHLS with the parameter scaling with density captured in the dilaton condensate $\la\chi\ra$ and in the FLFP approximation  could well be comparable to the high-order s$\chi$EFT calculation. Furthermore --- and this is not recognized in nuclear circles --- there is a magical element in HLS, as stressed in \cite{Komargodski,kanetal}, possibly connected with the Seiberg-duality between HLS and gluons of QCD.

Just to give an idea, we quote the predictions of a few thermodynamic properties of nuclear matter compared in parentheses with available empirical information: $n_0 = 0.161 (0.16\pm 0.01)$ fm$^{-3}$,  B.E. $= 16.7 (16.0\pm 1.0)$ MeV, $E_{sym} (n_0)=30.2 (31.7\pm 3.2)$ MeV,  $E_{sym} (2n_0)= 56.4 (46.9\pm 10.1; 40.2\pm 12.8)$ MeV, $L(n_0)=67.8 (58.9\pm 16;  58.7\pm 28.1)$ MeV, $K_0=250.0 (230\pm 20)$ MeV.

That the standard high-order $s\chi$EFT and the GnEFT with explicit hidden symmetries fare equally well at $n\sim n_0$ indicate that those symmetries are buried and not apparent in the EOS at that density. This does not mean all observables are opaque to them. Totally unrecognized in the past  is that the long-standing mystery of quenched $g_A$ in nuclei as seen in shell models  has a deep connection with how scale symmetry hidden in QCD emerges in nuclear correlations~\cite{gA}.

The superallowed matrix element of the Gamow-Teller transition from proton (neutron) to neutron (proton) on top of the Fermi sea calculated in GnEFT at the FLFP is given by the effective $g_A$
\be
g_A^{\rm Landau}/g_A = \Big(1-\frac 13 \Phi \tilde{F}_1^\pi\Big)^{-2}\label{gA}
\ee
where $g_A=1.27$ is the axial coupling constant for the free neutron decay, $\Phi=\la\chi\ra_\ast/\la\chi\ra_0$ and $\tilde{F}_1^\pi$ is the Landau $F_1$ parameter coming from the pion exchange. $\Phi$ gives the scaling in density of the dilaton condensate in medium and $\tilde{F}_1^\pi$ is exactly given at a given density by chiral symmetry. $\Phi$ can be related to the scaling of the in-medium pion decay constant $f_\pi$ so is known experimentally up to $\sim n_0$. The RHS of (\ref{gA}) turns out to be highly insensitive to density near $n_0$, so is nearly constant between $\frac 12 n_0$ and $n_0$: it comes out to be $\sim 0.8$.  Thus we get from medium-heavy nuclei to nuclear matter
\be
g_A^{\rm Landau}\approx 1.0.\label{gA1}
\ee
This is the famous quenching of $g_A=1.27$ to $g_A^{sm}\approx 1.0$ that has remained unexplained  since mid 1970's. As discussed in \cite{gA}, there can be corrections coming from conformal anomaly that enter at higher scale-chiral order which can be ``measured" in RIB experiments~\cite{sn100}. But the implication will not be significantly modified.

Equation (\ref{gA}) is justified in the large $N_c$ and large $\bar{N}$ limit in what corresponds to the Landau FLFP. It's derivation exploits soft dilaton theorems~\cite{crewther}.   As formulated it can be precisely equated to the superallowed  Gamow-Teller transition matrix element given in the extreme single-particle shell-model  of a doubly magic nuclei such as $^{100}$Sn~\cite{sn100}. At the matching scale to QCD,  the axial weak coupling to the nucleons is scale-invariant, hence the renormalization (\ref{gA}) can be considered entirely due to nuclear (many-body) interactions. Thus (\ref{gA1}) captures the influence of the scale symmetry  emergent from the (nuclear) interactions  which may or may not be directly connected to QCD. We can see this  also by going to high density $\gg n_0$. Starting with non-linear sigma model with constituent quarks it is shown that \cite{bira-dlfp} after the field redefinition, letting the dilaton mass $m_\chi$ go to zero with the conformal anomaly turned off --- and in the chiral limit --- leads to a linearized Lagrangian that satisfies various well-established sum rules, such as, among others, the Adler-Weisberger sum rule,  provided the singularities that appear as $m_\chi\to 0$ are suppressed by what are called ``dilaon-limit  fixed-point (DLFP)" constraints. The same limiting process to our $\chi$bHLS yields \cite{bira-dlfp}
\be
g_V&\to& g_A\to 1,\label{gagv}\\
f_\pi &\to& f_\chi\neq 0.\label{bira}
\ee
We denote the $g_A$ in this limit as $g_A^{DL}$.
The DLFP gives no constraint on the $\omega$-nucleon coupling into which we do not enter here. It is rather involved and has a crucial interplay in how the dilaton condensate becomes density-independent in the half-skyrmion phase leading to the PC velocity~\cite{MR-review}.

Now the important point  is that (\ref{gagv}) and (\ref{bira}) allow the DLFP to be identified with the IRFP~\cite{crewther}. In connection with the $g_A$ quenching, this provides the mechanism to make the emergent scale symmetry pervade from low density --- $g_A^{\rm Landau}\approx 1$ --- to  high density --- $g_A^{\rm DL}\approx 1$. We will argue this is also what happens  in compact stars.

\sect{Equation of State  of Compact Stars}
We now turn to the application of the GnEFT formalism to the structure of massive neutrons stars based on the standard TOV equation. We shall focus on the EOS of the baryonic matter, leaving out such basic issues as corrections to gravity, dark matters etc. Unless otherwise stated the role of leptons --- electrons, muons, neutrinos etc --- is included in the EOS. The results we present here are not new (as summarized in \cite{MR-review}), but  their implications and impacts on nuclear dynamics bring a totally new light and offer a paradigm change in nuclear physics.

Now to go to higher densities beyond $n_0$ in GnEFT for compact stars, what  we first need  is to fix the density at which the topology change takes place.
It turns out that the available astrophysics phenomenology does provide the range  $2\lsim n_{1/2}/n_0 < 4$~\cite{MR-review}. Here, we pick $n_{1/2}=2.5n_0$ for illustration.

Up to $n_{1/2}$, the same EOS that works well at $n_0$ is assumed to hold.  It is what comes at $n_{1/2}$ due to the topology change that plays the crucial role for the properties of compact stars. Among the various items listed in \cite{MR-review}, the most prominent are (a) the cusp in the symmetry energy $E_{sym}$ at $n_{1/2}$, (b) the VM with $m_\rho\to 0$ at $n_{vm}\gsim 25 n_0$, (c) the approach to the DLFP --- at or close to the putative IRFP --- at $n_{dl}\sim n_{vm}$  and (d) the hadron-quark continuity up to $n_{vm}$. The cusp in $E_{sym}$ leads  to the suppression of the $\rho$ tensor force, the most important for the symmetry energy, and triggers the effect (b). It effectively makes the EOS transit from soft-to-hard  in the EOS, thus accounting for massive $\gsim 2 M_\odot$ stars. The effects (b) and (c) --- together with the $\omega$ coupling to nucleons --- make the effective mass of the confined half-skyrmions go proportional to the dilaton decay constant $f_\chi \sim m_0$ which is independent of density. Thus {\it $f_\chi$ depends little on density in the half-skyrmion phase.}

The star properties obtained in GnEFT are found to be generally consistent with presently available observations~\cite{MR-review}. For  $n_{1/2}=2.5 n_0$,  the maximum star mass is found to be $M_{max}\sim 2.0 M_\odot$ with the central density $\sim 5n_0$. For a neutron star with mass $1.4M_{\odot}$, currently highly topical in connection with gravity-wave data,  we obtain the dimensionless tidal deformability $\Lambda_{1.4} \approx 656$ and the radius $R_{1.4} \approx 12.8$~km.

Our theory makes a basically different prediction from that of all  other ``standard" EFT available in the literature. And that is in the SV of stars $v_s$ and its impact on the structure of the core of massive stars.

\begin{figure}[htbp]
\begin{center}
\includegraphics[width=0.24\textwidth]{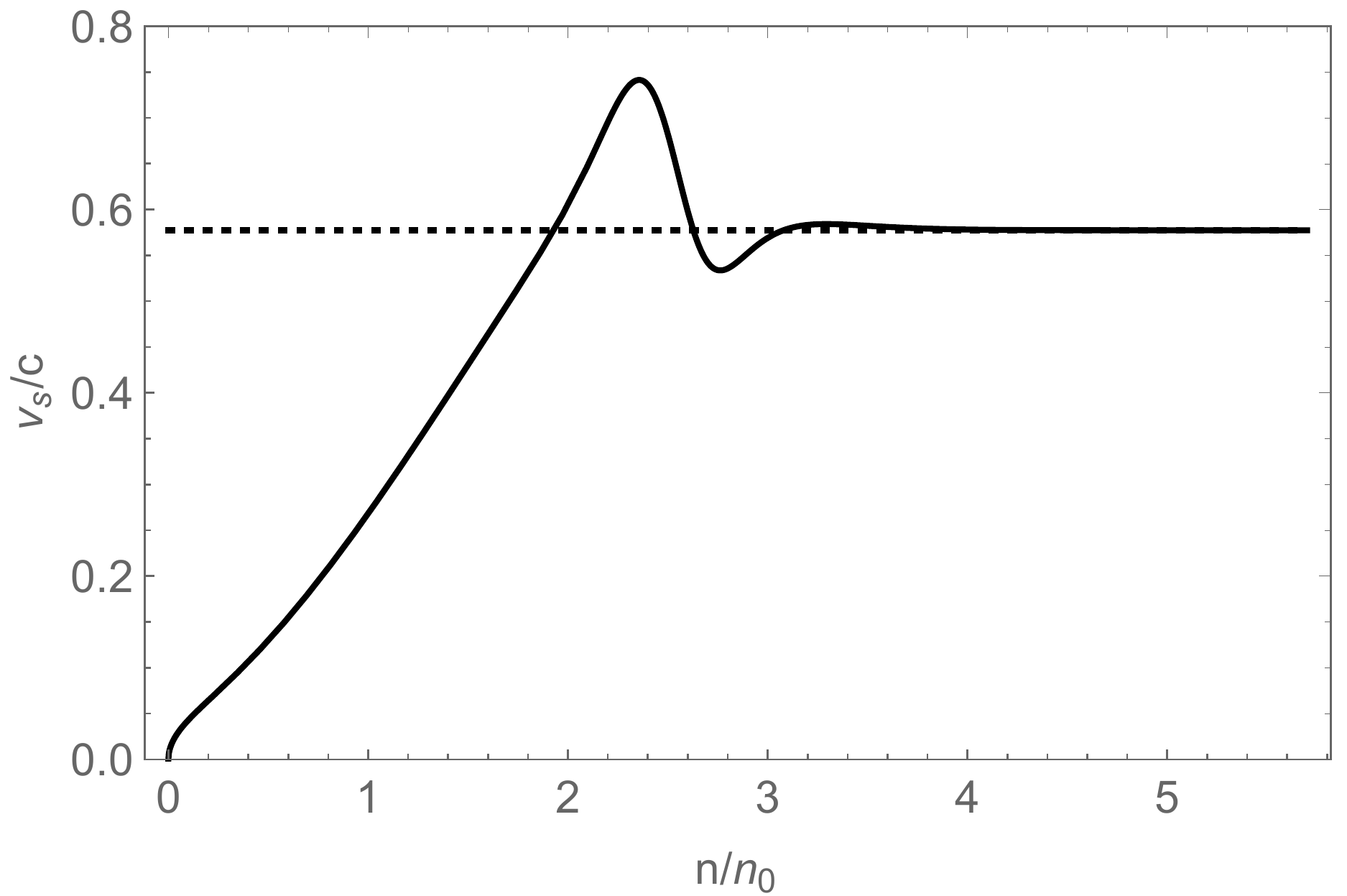}\includegraphics[width=0.24\textwidth]{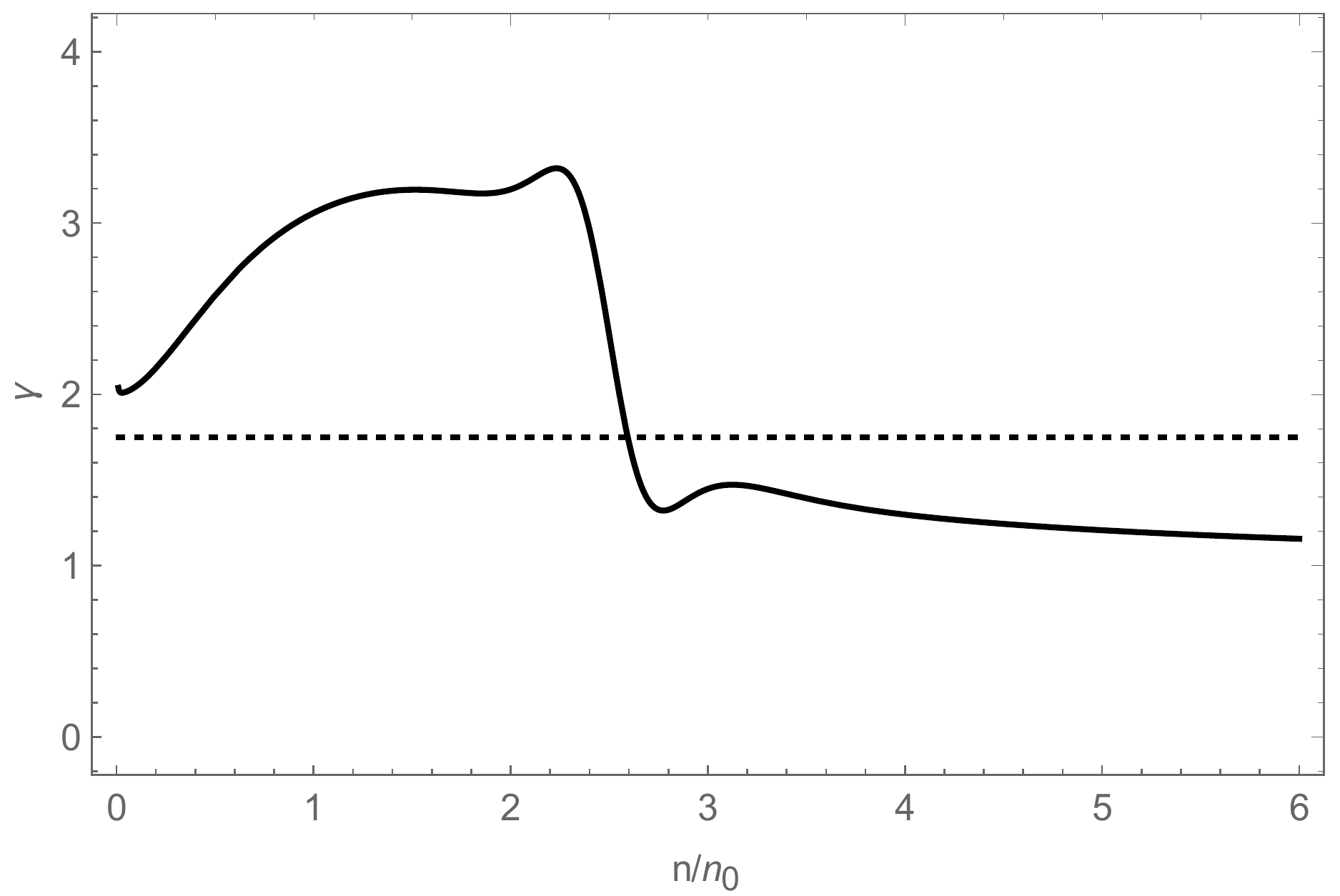}
\end{center}
\vskip -.5cm
\caption{Density dependence of the SV of stars $v_s$ (left panel) and the polytropic index $\gamma= {d\ln P}/{d\ln \epsilon}$ (right pannel) in neutron matter.}
\label{vs}
\end{figure}

\sect{Sound Velocity and Emergent Scale Symmetry} To see what'a going on with $v_s$, let's look at the trace of the energy-momentum tensor $\la\theta_\mu^\mu\ra$ which encodes the conformal anomaly. Suppose the dense matter is at the IRFP in the GD scheme (ignoring  quark masses),  then we will have $\la\theta_\mu^\mu\ra=0$ and hence the conformal SV $v_s^2/c^2=1/3$. But our star matter must be some distance away from the IRFP even if the quark masses are ignored as $m_\chi\neq 0$, so $\theta^\mu_\mu$ cannot be zero. Yet  a surprise in the prediction of GnEFT was that the SV came out to be the {\it same} as the conformal SV as shown in Fig.~\ref{vs} (left panel) for $n\gsim3n_0$ for the assumed VM density (which we identify with the deconfinment density as stated below) $n_{vm}\sim n_{dl} \gsim 25 n_0$~\cite{PKLMR}. Were the $n_{vm}$ much lower, say, $\sim 6 n_0$, then $v_s^2/c^2$ would increase continuously,  passing 1/3 at $\sim 3 n_0$ and reaching  $\gsim 1/2$ at $\sim 9 n_0$~\cite{PKLR}. Thus the possible ``deconfinement density" seems to play an important role for the SV whereas  there are no appreciable  differences in other star properties.

The calculations were done in the $V_{lowk}$ RG formalism which includes certain $1/\bar{N}$ corrections, going beyond the Landau FLFP approximation. How the striking result given above comes about in our EFT can be understood in an extremely simple way. In the mean field of  GnEFT, the VEV of the trace of energy momentum tensor takes the simple form $
\la\theta^\mu_\mu\ra=4V(\la\chi\ra) -\la\chi\ra\frac{\del V(\chi)}{\del\chi}|_{\chi=\la\chi\ra}$
where $V(\chi)$ is the dilaton potential where the conformal anomaly effect is included. As noted above, $\la\chi\ra\propto f_\chi$ is density-independent in the range of density involved in the interior of compact stars, so $
\frac{\partial}{\partial n}\langle \theta_\mu^\mu \rangle = \frac{\partial \epsilon(n)}{\partial n}(1-3v_s^2/c^2)=0.$ Now if $\frac{\partial \epsilon(n)}{\partial n}\neq 0$ (i.e., no Lee-Wick state), then we arrive at $1-3v_s^2/c^2=0$. Since our system is necessarily some distance away from the IRFP to account for the non-zero dilaton mass, $\la\theta^\mu_\mu\ra$ cannot be zero. The $v_s$ cannot be conformal. Thus though numerically the same, we call this  ``pseudo-conformal"
\be
v_{pcs}^2/c^2\approx 1/3.\label{pcs}
\ee
It is of course approximate because both the quark mass terms and the deviation from the IRFP --- which must be present in the EOS --- are ignored.
\begin{figure}[htbp]
\begin{center}
\vskip 0.3cm
\includegraphics[width=0.32\textwidth]{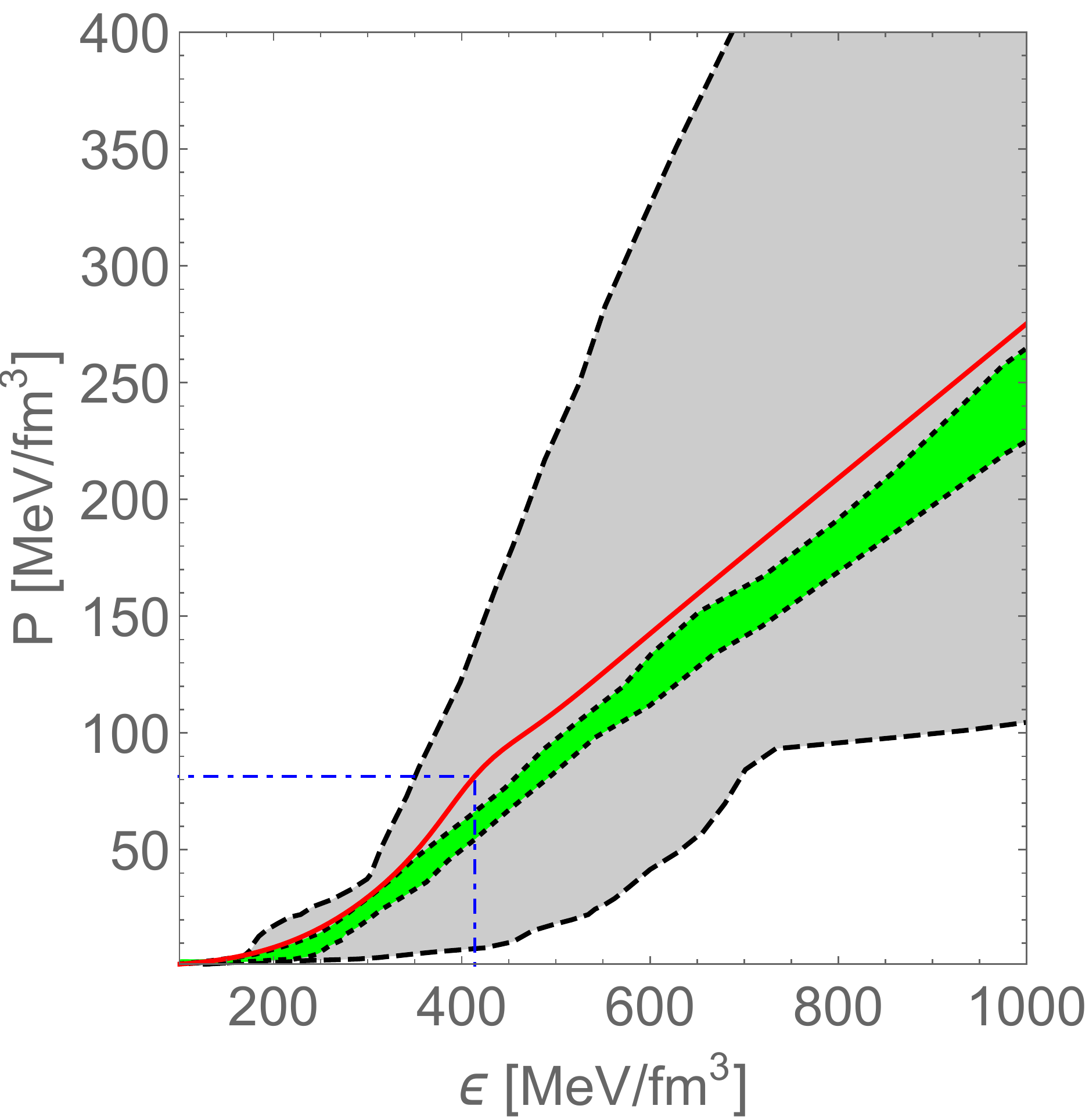}
\end{center}
\vskip -0.5cm
\caption{Comparison of $(P/\epsilon)$  between the PCM  velocity and the band generated with the
SV interpolation method used in~\cite{evidence}. The gray band is  from the causality and the green band from the conformality. The red line is the PCM prediction. The dash-dotted line indicates the location of the topology change.}
\label{fig:EoS}
\end{figure}

\sect{Confronting Nature}
This prediction can be confronted with a recent analysis that combines astrophysical observations and model independent theoretical {\it ab initio} calculations~\cite{evidence}. Based on the observation that, in the core of the maximally massive stars, $v_s$ approaches the conformal limit $v_s/c \to 1/\sqrt{3}$ and the polytropic index takes the value $\gamma < 1.75$ --- the value close to the minimal one obtained in hadronic models --- Annala et al. arrive at the conclusion that the core of the massive stars is populated by  ``deconfined" quarks.

It is perhaps surprising that the predicted pseudo-conformal speed (\ref{pcs}) sets in {\it precociously} at $\sim 3n_0$ and stays constant in the interior of the star.  Microscopic descriptions such as the quarkyonic model~\cite{quarkyonic} typically exhibit more complex structures at the putative hadron-quark transition density. We think the simpler structure in our description  is due to perhaps too drastic suppression of higher-order $1/\bar{N}$ terms in the half-skyrmion phase. Similarly in the right panel, one sees the polytropic index $\gamma$  drops, again rapidly, below 1.75 at $\sim 3n_0$ and approaches 1 at $n\gsim 6n_0$.

Finally --- and most importantly --- we compare in Fig.~\ref{fig:EoS} our prediction for $P/\epsilon$ with the conformality band obtained by the SV interpolation method ~\cite{evidence}. We see that it is close to, and parallel with, the conformality band, but most significantly, it lies outside of this band.

The predicted results of GnEFT as a whole resemble  the ``deconfined" quark structure of \cite{evidence}. There are, however, basic differences between the two. First of all,  in our theory, conformality is broken, though perhaps only slightly at high density, in the system. This could be related to the deviation of $g_A^{\rm Landau}$ from the experimental value of the quenching in $^{100}$Sn observed~\cite{gA}. There can also be fluctuations around $v_{pcs}^2/c^2=1/3$ coming from the effects by the anomalous dimension $\beta^\prime$. This effect can be seen in Fig.~{\ref{fig:EoS} where the PCM prediction deviates slightly from the ``would -be" conformal band. Most important of all, the confined half-skyrmion fermion in the half-skyrmion phase is {\it not} deconfined. It is a quasiparticle of fractional baryon charge, neither purely baryonic nor purely quarkonic. In fact it can be anyonic lying on a (2+1) dimensional sheet~\cite{D}. What it represents is a manifestation of an emergent scale symmetry pervading at low density as in $g_A^{\rm Landau}$ and  in $g_A^{DL}$ at high density in the vicinity of DLFP. We suggest to classify the precocious pseudo-conformal SV in the same class of emerging scale symmetry in action in nuclear processes. In fact it is in line with how conformal symmetry permeates from the unitarity limit in light nuclei~\cite{bira-unitality} to the symmetry energy near $n_0$~\cite{lattimer-unitarity} and more.

\sect{Conclusion} We have suggested that quark-like degrees of freedom, if observed in the interior of massive neutron stars, can be interpreted as confined quasi-particles of fractional baryon charges in consistency with hadron-quark continuity. Such fractionally-charged objects are inevitable by topology at high densities~\cite{D}. The mechanism in action is the emergence of conformal (or scale) symmetry, coming not necessarily from QCD proper, but from strongly-correlated nuclear interactions, which could permeate, either hidden or exposed, in baryonic matter from low density to high density. In this scheme, true deconfinement is to set in as mentioned above at much higher densities, say, $\gsim 25 n_0$,  than relevant to compact stars when the VM fixed point and/or DLFP are reached, possibly with the phase transition from a Higgs mode to a topological mode~\cite{kanetal}.

\sect{Acknowledgments}
We are grateful for helpful correspondence from Jim Lattimer and Aleksi Vuorinen on their recent publications. The work of YLM was supported in part by the National Science Foundation of China (NSFC) under Grant No. 11875147 and 11475071.



\begin{thebibliography}{50}

\bi{baymetal}  G.~Baym, S.~Furusawa, T.~Hatsuda, T.~Kojo and H.~Togashi,
  ``New neutron star equation of state with quark-hadron crossover,''
  Astrophys.\ J.\  {\bf 885}, 42 (2019).

\bi{quarkyonic} L.~McLerran and S.~Reddy,
  ``Quarkyonic matter and neutron stars,''
  Phys.\ Rev.\ Lett.\  {\bf 122}, no. 12, 122701 (2019);   K.~S.~Jeong, L.~McLerran and S.~Sen,
  ``Dynamically generated momentum space shell structure of quarkyonic matter via an excluded volume model,''
  Phys.\ Rev.\ C {\bf 101}, no. 3, 035201 (2020); T.~Zhao and J.~M.~Lattimer,
  ``Quarkyonic matter equation of state in beta-equilibrium,''
  arXiv:2004.08293 [astro-ph.HE].


\bibitem{Holt:2014hma}
J.~W.~Holt, M.~Rho and W.~Weise,
``Chiral symmetry and effective field theories for hadronic, nuclear and stellar matter,''
Phys. Rept. \textbf{621}, 2-75 (2016).







\bi{collins-perry} J.~C.~Collins and M.~J.~Perry,
  ``Superdense matter: Neutrons or asymptotically free quarks?,''
  Phys.\ Rev.\ Lett.\  {\bf 34}, 1353 (1975).

 \bi{yamawaki} M.~Bando, T.~Kugo, S.~Uehara, K.~Yamawaki and T.~Yanagida,
  ``Is $\rho$ meson a dynamical gauge boson of hidden local yymmetry?,''
  Phys.\ Rev.\ Lett.\  {\bf 54}, 1215 (1985).
\bi{suzuki} M.~Suzuki,
  ``Inevitable emergence of composite gauge bosons,''
  Phys.\ Rev.\ D {\bf 96}, no. 6, 065010 (2017).

 \bi{Komargodski} Z.~Komargodski,
  ``Vector mesons and an interpretation of Seiberg duality,'' JHEP {\bf 1102}, 019 (2011).


\bibitem{kanetal} N.~Kan, R.~Kitano, S.~Yankielowicz and R.~Yokokura,
  ``From 3d dualities to hadron physics,'' arXiv:1909.04082 [hep-th].

\bibitem{karasik} A.~Karasik,
  ``Skyrmions, quantum Hall droplets, and one current to rule them all,''
  arXiv:2003.07893 [hep-th].



\bi{VM}   M.~Harada and K.~Yamawaki,
  ``Hidden local symmetry at loop: A New perspective of composite gauge boson and chiral phase transition,''
  Phys.\ Rept.\  {\bf 381}, 1 (2003).

 \bi{D} Y.~L.~Ma and M.~Rho,
  ``Dichotomy of baryons as quantum Hall droplets and skyrmions In compact-star matter,''
  arXiv:2009.09219 [nucl-th].

\bi{crewther} R.~J.~Crewther,
  ``Genuine dilatons in gauge theories,''
  Universe {\bf 6}, no. 7, 96 (2020);
  R.~J.~Crewther and L.~C.~Tunstall,
  ``$\Delta I=1/2$ rule for kaon decays derived from QCD infrared fixed point,''
  Phys.\ Rev.\ D {\bf 91}, no. 3, 034016 (2015)

  \bi{MR-review}
Y.~L.~Ma and M.~Rho,
``Towards the hadron-quark continuity via a topology change in compact stars,''
Prog. Part. Nucl. Phys. \textbf{113}, 103791 (2020).

\bi{shankar} R.~Shankar,
  ``Renormalization group approach to interacting fermions,''
  Rev.\ Mod.\ Phys.\  {\bf 66}, 129 (1994).

\bi{MR91} M.~Rho,
  ``Exchange currents from chiral Lagrangians,''
  Phys.\ Rev.\ Lett.\  {\bf 66}, 1275 (1991); B.~Friman and M.~Rho,
  ``From chiral Lagrangians to Landau Fermi liquid theory of nuclear matter,''
  Nucl.\ Phys.\ A {\bf 606}, 303 (1996).

\bi{DQCP}  T.~Senthil, A.~Vishwanath, L.~Balents, S.~Sachdev and M.~P.~A.~Fisher,
  ``Deconfined quantum critical points,''
  Science {\bf 303}, no. 5663, 1490 (2004).

 \bi{cho}
P.~Zhang, K.~Kimm, L.~Zou and Y.~M.~Cho,
  ``Re-interpretation of Skyrme theory: New topological structures,''
  arXiv:1704.05975 [hep-th];
 W.~S.~Bae, Y.~M.~Cho and B.~S.~Park,
  ``Reinterpretation of Skyrme theory,''
  Int.\ J.\ Mod.\ Phys.\ A {\bf 23}, 267 (2008);
S.~B.~Gudnason and M.~Nitta,
  ``Fractional skyrmions and their molecules,''
  Phys.\ Rev.\ D {\bf 91}, no. 8, 085040 (2015).

\bi{gA} Y.~L.~Ma and M.~Rho,
  ``The quenched ${g_A}$ in nuclei and emergent scale symmetry in baryonic matter,''
  Phys.\ Rev.\ Lett.\  {\bf 125}, no. 14, 142501 (2020).

 \bi{sn100}
 C.B. Henke {\it et al.},
``Superallowed Gamow-Teller decay of the doubly magic nucleus $^{100}$Sn,"
 Nature {\bf 486}, 341 (2012).


  \bi{bira-dlfp}  S.~R.~Beane and U.~van Kolck,
  ``The dilated chiral quark model,'' Phys.\ Lett.\ B {\bf 328}, 137 (1994).

 \bi{PKLMR} W.~G.~Paeng, T.~T.~S.~Kuo, H.~K.~Lee, Y.~L.~Ma and M.~Rho,
  ``Scale-invariant hidden local symmetry, topology change, and dense baryonic matter. II.,''
  Phys.\ Rev.\ D {\bf 96}, no. 1, 014031 (2017).

  \bibitem{PKLR}
  W.~G.~Paeng, T.~T.~S.~Kuo, H.~K.~Lee and M.~Rho,
  ``Scale-Invariant hidden local symmetry, topology change and dense baryonic matter,''
  Phys.\ Rev.\ C {\bf 93}, no. 5, 055203 (2016).


  \bi{evidence}  E.~Annala, T.~Gorda, A.~Kurkela, J.~N\"attil\"a and A.~Vuorinen,
  ``Evidence for quark-matter cores in massive neutron stars,''
  Nature Phys.\  (2020) doi:10.1038/s41567-020-0914-9
  [arXiv:1903.09121 [astro-ph.HE]].

 \bi{bira-unitality} S.~K\"onig, H.~W.~Griesshammer, H.~W.~Hammer and U.~van Kolck,
  `Nuclear Physics around the unitarity limit,''
  Phys.\ Rev.\ Lett.\  {\bf 118}, no. 20, 202501 (2017).

  \bi{lattimer-unitarity}  I.~Tews, J.~M.~Lattimer, A.~Ohnishi and E.~E.~Kolomeitsev,
  ``Symmetry parameter constraints from a lower bound on neutron-matter energy,''
  Astrophys.\ J.\  {\bf 848}, no. 2, 105 (2017).






\end{thebibliography}
\end{document}